\documentclass[twocolumn]{aastex63}  
\usepackage{comment}
\usepackage{url}

\received{March 18, 2021}
\accepted{June 8, 2021}
\submitjournal{ApJ}

\shorttitle{Energy conversion rate of a transient brightening estimated by Hinode/EIS.}
\shortauthors{Kawai and Imada}

\begin{document}

\title{Energy conversion rate of an active region transient brightening estimated by a spectroscopic observation of Hinode}

\correspondingauthor{Toshiki Kawai}
\email{t.kawai@isee.nagoya-u.ac.jp}

\author[0000-0003-0521-6364]{Toshiki Kawai}
\affiliation{Institute for Space-Earth Environmental Research, Nagoya University, \\
Furo-cho, Chikusa-ku, Nagoya, Japan}

\author{Shinsuke Imada}
\affiliation{Institute for Space-Earth Environmental Research, Nagoya University, \\
Furo-cho, Chikusa-ku, Nagoya, Japan}



\begin{abstract}
We statistically estimate the conversion rate of the energy released during an active-region transient brightening to Doppler motion and thermal and non-thermal energies. We used two types of datasets for the energy estimation and detection of transient brightenings. One includes spectroscopic images of Fe\,{\sc xiv}, Fe\,{\sc xv}, and Fe\,{\sc xvi} lines observed by the Hinode/EUV Imaging Spectrometer. The other includes images obtained from the 211~\AA~ channel of the Solar Dynamics Observatory/Atmospheric Imaging Assembly (AIA). The observed active region was NOAA 11890 on November 09, 2013, and the day after that. As a result, the released Doppler motion and non-thermal energies were found to be approximately 0.1 -- 1\% and 10 -- 100\%  of the change in the amount of thermal energy in each enhancement, respectively. Using this conversion rate, we estimated the contribution of the total energy flux of AIA transient brightenings to the active region heating to be at most 2\% of the conduction and radiative losses. 
\end{abstract}



\section{Introduction} \label{sec:intro}
The mechanism of heating of the solar corona is still unknown, which is one of the most important problems in solar physics. 
There are two primary models to explain this mechanism: small-scale magnetic reconnection and wave dissipation. 
In the former model, the corona is heated by small-scale impulsive heating events, the so-called nanoflares ($E \simeq 10^{24}~\mathrm{erg}$), which are associated with the magnetic reconnections \citep[{\it e.g.,\rm}][]{Parker1988}. 
In the latter model, the corona is heated by the dissipation of Alfv\'{e}n waves from the surface, which are generated by convection \citep[{\it e.g.,\rm}][]{Goldstein1978, Hollweg1982}.
Recently, considerable progress has been made through theoretical and observational studies for both models; however, a definitive solution to this problem has yet to be achieved \citep[{\it e.g.,\rm}][]{Mandrini2000, Klimchuk2006, Parnell2012, Klimchuk2015}.
From the perspective of the nanoflare heating model, it is crucial to quantify the contribution of small-scale flares in the heating of the corona accurately. 

\citet{Parker1988} proposed that nanoflares are magnetic reconnections between the coronal magnetic fields tangled by the foot point motions due to the convection, as described in \citet{Parker1983}. 
The first observation of a microflare was carried by hard X-ray balloon observations in 1980 \citep{Lin1984}. 
\citet{Shimizu1992, Shimizu1994} observed many small explosive events which occur in active regions using the {\it Soft X-ray Telescope} \citep[SXT:][]{Tsuneta1991} onboard the Yohkoh satellite \citep{Ogawara1991}.
To validate the nanoflare heating model, it is essential to reveal the occurrence frequency distribution of the flares as a function of energy. 
The distribution is known to follow a power law, as given by the following equation:
\begin{equation}
	\frac{dN}{dE}=AE^{-\alpha}
	\label{eq:alpha}
\end{equation}
where $N$, $E$, $A$, and $\alpha$ are the number of events, energy of each event, power law constant, and power-law index, respectively \citep{Hudson1991}.
The total energy released by all flares $P$ can be calculated as follows: 
\begin{equation}
	P=\int_{E_{\mathrm{min}}}^{E_{\mathrm{max}}}\frac{dN}{dE}EdE=\frac{A}{-\alpha+2}\left(E_{\mathrm{max}}^{-\alpha+2}-E_{\mathrm{min}}^{-\alpha+2}\right)
\end{equation}
Therefore, small-scale flares dominantly heat the corona when $\alpha$ is greater than two, and $E_{\mathrm{min}}$ is sufficiently small.
The frequency distributions of smaller flares have been estimated using various methods and instruments for a few decades. 
In some studies, $\alpha$ was found to be greater than the threshold 2 \citep[{\it e.g., \rm}][]{Parnell2000, Benz2002}, which suggests that the frequency distribution at smaller energies can be broken to heat the corona sufficiently, whereas other studies found that $\alpha$ is smaller \citep[e.g.,\rm][]{Shimizu1995, Aschwanden2000, Tajfirouze2016, Jess2019}.
This uncertainty might be caused by the failure to detect the smallest events or energy estimation due to instrumental limitations or wrong methods. 
One of the problems in these studies is that their analysis is based on EUV and X-ray thermal energies but not on kinetic or non-thermal energies. 
\citet{Kawai2021} derived the contribution of small-scale flares using one-dimensional loop simulation and a genetic algorithm. 
One of the strengths of their study is that the method can consider the distribution of released energy to thermal and non-thermal energy using simulations.
They suggested that $\alpha$ is greater than two in the energy range of $10^{26} \lesssim E \lesssim 10^{27.5}$, whereas $\alpha$ is less than two for smaller flares. 

Spectroscopic observations have successfully achieved significant progress in coronal plasma physics for example, flares \citep{Imada2013, Imada2014, Polito2018b}, jets \citep{Matsui2012, Young2014, Kawai2019}, dimmings \citep{Imada2011b}, and coronal loops \citep{Schmelz2001, Warren2008}. 
One of the strengths of spectral observations is the ability to derive a line-of-site plasma velocity for each ion species. 
\citet{Testa2014} reported the rapid variation of intensity and velocity by the {\it Interface Region Imaging Spectrograph} \citep[IRIS:][]{DePontieu2014} launched in 2013.
IRIS provides imaging and spectral observations of the chromosphere and transition region at high spatial, temporal, and wavelength resolutions. 
According to the comparison with a simulation, this event is caused by a non-thermal electron beam accelerated by a nanoflare that has an energy of $10^{25}~\mathrm{erg}$. 
Using one-dimensional coronal loop simulation, \citet{Polito2018} suggested that up to $20~\mathrm{km~s^{-1}}$ of the Doppler blueshift can be seen in the {\it IRIS} Si\,{\sc iv} line when non-thermal electron beams are produced even by $10^{24}~\mathrm{erg}$ events. 
\citet{Brooks2016} reported that the mean value of non-thermal velocity in the non-flaring active region is approximately $18~\mathrm{km/s}$ from observations of the {\it EUV Imaging Spectrometer}~\citep[EIS:][]{Culhane2007} on board Hinode~\citep{Kosugi2007}. 
This non-thermal velocity is much smaller than that expected from high-temperature reconnection jets in the nanoflare heating model. 

In this study, we estimate Doppler motion and non-thermal energies of nanoflares and the changes in the amount of thermal energy during them to compare the energy balance between them. 
Section~\ref{sec:obs} shows observational data of an active region obtained from {\it Hinode}/EIS  and the {\it Atmospheric Imaging Assembly} \citep[AIA:][]{Lemen2012} on board the {\it Solar Dynamics Observatory}~\citep[SDO:][]{Pesnell2012}. 
Sections ~\ref{sec:detect} and~\ref{sec:energy}  present the method of detecting events and calculating their Doppler motion and non-thermal and thermal energies. 
Section~\ref{sec:result} provides our results, and we discuss the energy balance between them and the contributions to active region heating in Section~\ref{sec:summary}. 

\section{Data and Observations} \label{sec:obs}
\begin{figure}[tp]
	\centering
	\includegraphics[width=1\linewidth]{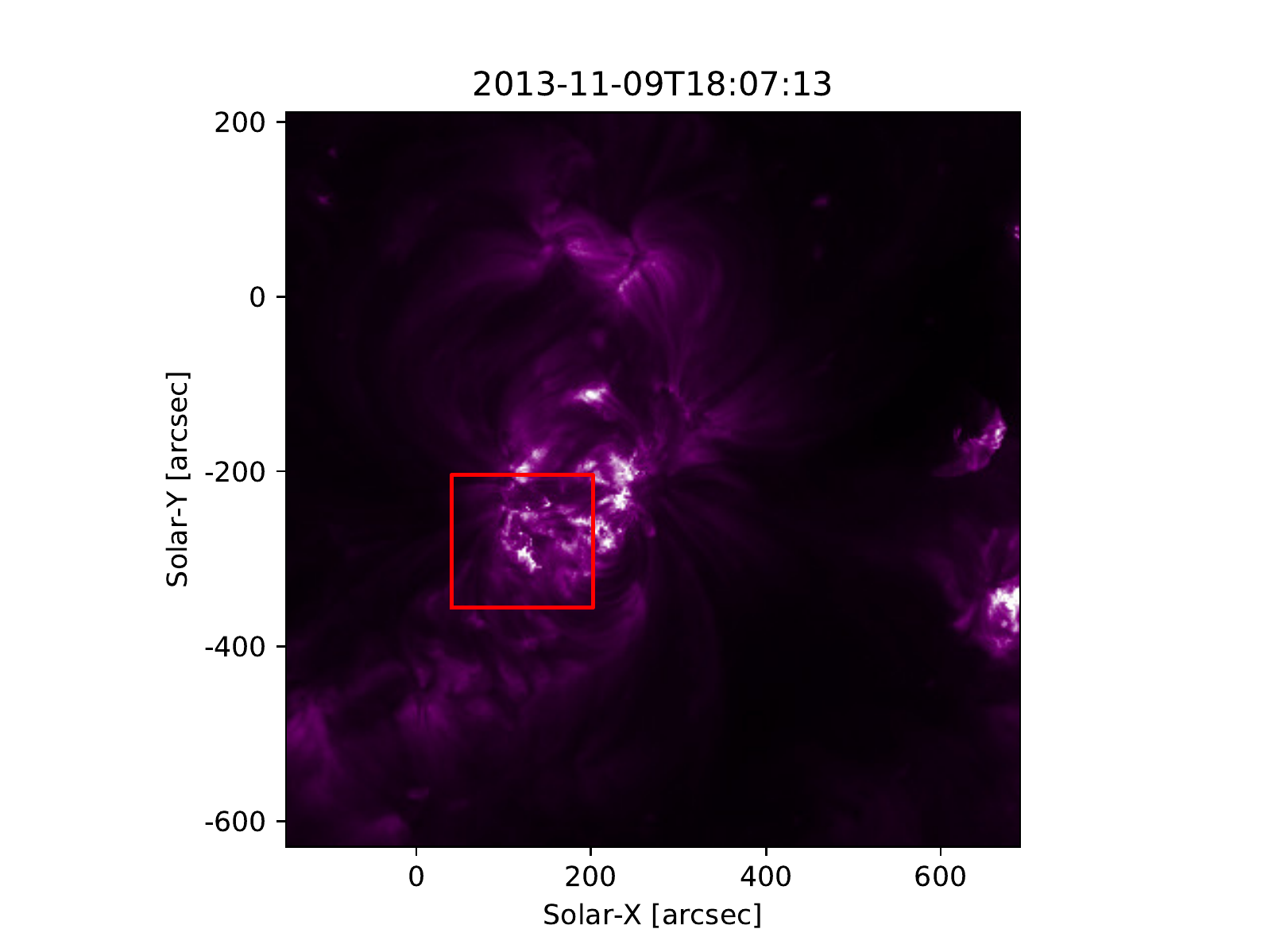}
	\caption{An example of maps of active region 11890 obtained from SDO/AIA 211~\AA. The coordinates of this active region are approximately (Solar-X, Solar-Y) = (100\arcsec, -200\arcsec). The red square indicates the field of view of the EIS at that time. }
	\label{fig:211}
\end{figure}
\begin{figure*}[tp]
	\centering
	\includegraphics[width=1\linewidth]{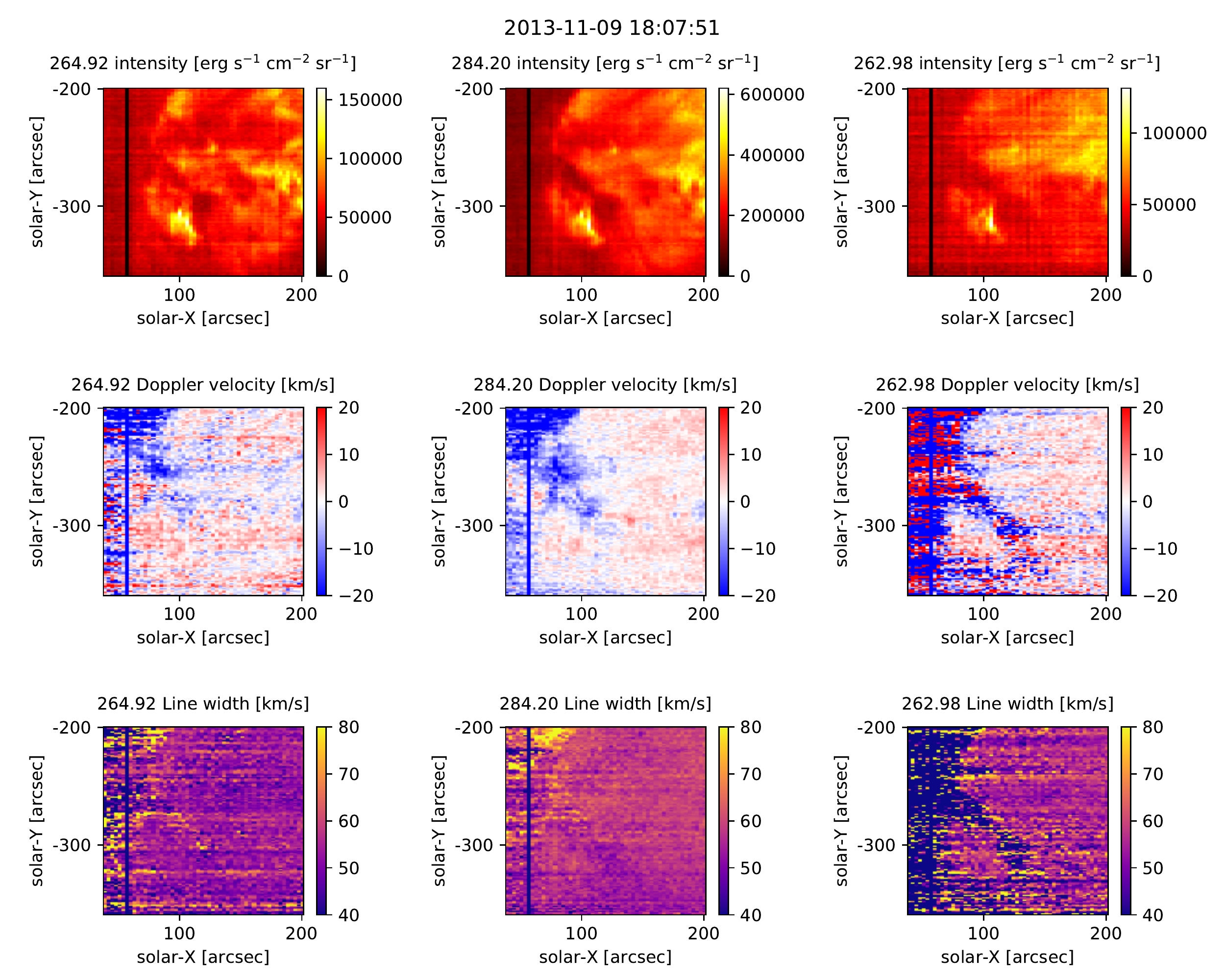}
	\caption{An example of maps of active region 11890 obtained from Hinode/EIS Fe\,{\sc xiv} ({\it left}), Fe\,{\sc xv} ({\it center}), and Fe\,{\sc xvi} ({\it right}). {\it top}, {\it middle}, and {\it bottom} panels present maps of the intensity, Doppler velocity, and line width, respectively. }
	\label{fig:eismap}
\end{figure*}

\begin{table}[tpb!]
	\centering
	\caption{Line list of Hinode/EIS study ID 485}
	  \begin{tabular}{rrr}
	  Ion species & Wavelength~[$\AA$] & $\log{\mathrm{(Formation~}T\mathrm{~[K])}}$ \\ \hline
Fe\,{\sc x} & 184.54 & 6.1 \\
Fe\,{\sc viii} & 185.25 & 5.7 \\
Fe\,{\sc xii} & 186.88 & 6.2\\
Fe\,{\sc xxiv} & 192.03 & 7.2\\
Ca\,{\sc xvii} & 192.83 & 6.8\\
Fe\,{\sc xii} & 195.12 & 6.2\\
Fe\,{\sc xvii} & 255.05 & 6.8\\
Fe\,{\sc xvi} & 262.98 & 6.8\\
Fe\,{\sc xxiii} & 263.69 & 7.2\\
Fe\,{\sc xiv} & 264.92 & 6.3\\
Fe\,{\sc xiv} & 274.37 & 6.3\\
Fe\,{\sc xv} & 284.20 & 6.4\\
	   \end{tabular}
	\label{tab:wvl}
\end{table}

In this study, we used two series of EUV images of the NOAA active region 11890 observed from 18:07 UT on November 09, 2013, to 14:20 UT on the next day. 
The first series includes spectroscopic images obtained from Hinode/EIS, and the other includes the EUV images observed by the SDO/AIA.  
Figure~\ref{fig:211} presents a snapshot of the active region at the beginning of the observation duration. 
The red square indicates the field of view of the EIS at that time. 
The coordinates of this active region are approximately (Solar-X, Solar-Y) = (100\arcsec, -200\arcsec) at the beginning of the observation. 
This active region causes GOES X1.1 flare, which begins at 05:08 UT on November 10 and continues for 10 min according to the Hinode Flare Catalog \citep{Watanabe2012}. 
This large brightening makes it difficult to detect small-scale events; however, as described below, we did not use the data around this time. 
There are no more flares that are greater than the GOES C-class.

We used the data series of the Hinode EIS Study ID 485, which was designed for flare observations. 
This study included 12 wavelength windows, as described in Table~\ref{tab:wvl}. 
The formation temperatures were obtained from the line list of the atomic database CHIANTI version 9.0.1 \citep{Dere2019}. 
We selected the Fe\,{\sc xiv}, Fe\,{\sc xv}, and Fe\,{\sc xvi} lines for our analysis because they have suitable formation temperatures for small-scale heating events and have less noise.  
Figure~\ref{fig:eismap} presents the first snapshot of the observations obtained by the Fe\,{\sc xiv} ({\it left}), Fe\,{\sc xv} ({\it center}), and Fe\,{\sc xvi} ({\it right}). The {\it top}, {\it middle}, and {\it bottom} panels show maps of the intensity, Doppler velocity, and line width, respectively. 
The method for calculating the Doppler velocity and line width is described in the next section. 
In cases data are lacking, as shown by the vertical line around the solar-X $\simeq 50~\mathrm{\arcsec}$ in Figure~\ref{fig:eismap} and Fe\,{\sc xvi}, because the data are relatively noisy. 
The field of view (FOV) of this study was $162~\mathrm{\arcsec}$ for solar-X and  $152~\mathrm{\arcsec}$ for solar Y. 
The number of pointing positions, slit width, scan step size, exposure time, and exposure delay were 54, 2\arcsec, 3\arcsec, $3~\mathrm{s}$, and $0~\mathrm{s}$, respectively. 
The resolution along the solar-Y is 1\arcsec. 
Time to complete each raster scan for making an image is approximately 254~s. 
Level 0 data can be retrieved from \footnote[1]{\url{http://solarb.mssl.ucl.ac.uk/SolarB/index.jsp}}Hinode/EIS website. 
EIS data from the raster were processed using procedures provided in SolarSoftWare \citep[SSW:][]{Freeland1998} to correct for flat fields, dark current, cosmic rays, hot pixels, and slit tilts. 
Owing to the temperature variation of the telescope, there was an orbital change in the line center, which caused an artificial Doppler shift of at most $\pm 20~\mathrm{km~s^{-1}}$. 
This error was corrected using the housekeeping data by \citet{Kamio2010}. 
Except for suspensions over 300 s, the observation duration was 30,831~s in total. 
We used 130 datasets for analysis in this study. 

\begin{figure}[tp]
	\centering
	\includegraphics[width=1\linewidth]{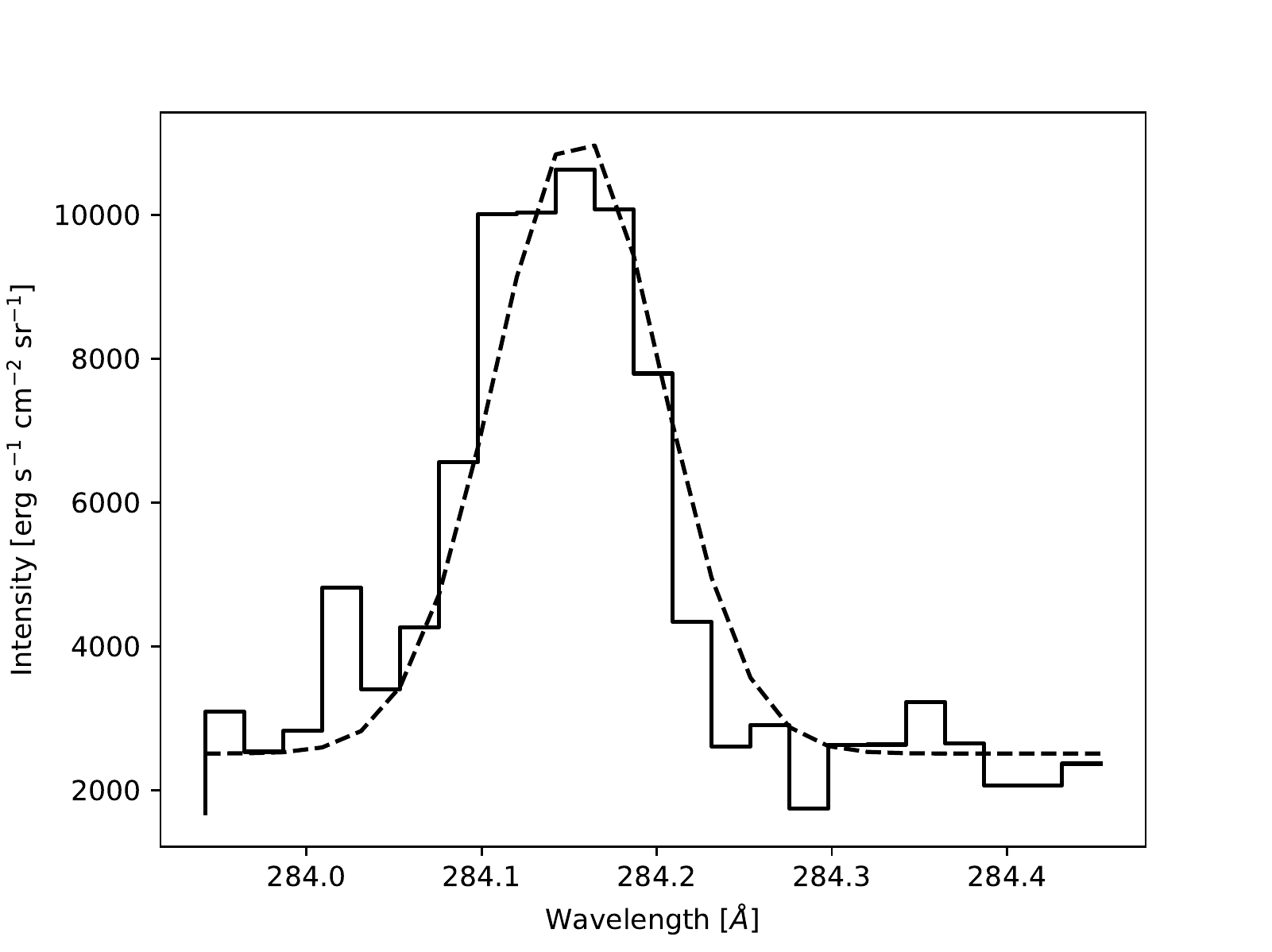}
	\caption{An example of the observed spectra obtained from the EIS Fe\,{\sc xv} window. The dashed line indicates the Gaussian fitted curve.}
	\label{fig:spc}
\end{figure}
Owing to the large amount of spectral data, each observed spectrum was automatically fitted by a single Gaussian function, as follows:
\begin{equation}
f(\lambda) = A_0 \exp \left[ -\frac{1}{2}\left(\frac{\lambda-A_1}{A_2}\right)^2 \right] + A_3
\label{eq:gauss}
\end{equation}
where Gaussian coefficients from $A_0$ to $A_3$ represent the height, center, standard deviation, and constant bias of the Gaussian, respectively. 
After fitting, the mean Doppler velocity and non-thermal velocity of the plasma can be retrieved from the spectrum. 
First, we defined the line-of-site velocity in each pixel $v_{\mathrm{Dop}}^\prime$ as follows:
\begin{equation}
v_{\mathrm{Dop}}^\prime = \frac{A_1 - \lambda_0}{\lambda_0}c
\end{equation}
where $\lambda$ and $c$ are the target wavelength and speed, respectively. 
Then, we defined the Doppler velocity by calibrating $v_{\mathrm{Dop}}^\prime$ using their median in each image as
\begin{equation}
v_{\mathrm{Dop}} = v_{\mathrm{Dop}}^\prime - \mathrm{Median}\left(v_{\mathrm{Dop}}^\prime\right). 
\end{equation}

The line width of each spectrum, which indicates the non-thermal velocity, can be derived as follows:
\begin{equation}
\delta \lambda = \frac{1}{\sqrt{4 \ln 2}}\frac{c}{\lambda_0} A_2 . 
\end{equation}
However, this line width is generally estimated to be larger than the non-thermal velocity of the plasma because $w$ includes line broadening due to instrumental width. 
Therefore, we estimated the non-thermal velocity $v_{\mathrm{nth}}$ according to \citet{Brooks2016}.
\begin{equation}
v_{\mathrm{nth}} = \sqrt{\delta \lambda^2 - \left(\frac{2k_BT_i}{m_i} + \sigma_I^2 \right)}
\label{eq:nth_vel}
\end{equation}
where $\lambda_0$, $k_B$, $T_i$, $m_i$, and $\sigma_I$ represent the line centroid, Boltzmann's constant, ion temperature, mass, and instrumental width, respectively.
The line centroid, $\lambda_0$, and ion temperature, $T_i$, are defined based on the values on the line list of the CHIANTI database. 
The instrumental width was measured in the laboratory before the launch of the Hinode by \citet{Korendyke2006}.
Young P. R. derived the Y-axis variation of the instrumental width from off-limb quiet sun observations. 
The slit width, which includes their correlation, is provided in SSW as a routine called \verb|eis_slit_width|. 
We used values in this routine for $\sigma_I$ similar to \citet{Brooks2016}. 
The mean ion mass value we used was $9.27 \times 10^{-23}$ g, according to the mass of the iron ion.

An example of the observed spectra (solid) and Gaussian fitted curve (dashed line) is shown in Figure~\ref{fig:spc}.
This spectrum is obtained from the EIS Fe\,{\sc xv} window. 
In this case, the Doppler and non-thermal velocities are $-26~\mathrm{km~s^{-1}}$ and $42~\mathrm{km~s^{-1}}$, respectively. 

The time resolution of EIS raster images was slightly long to analyze small-scale brightenings that have time scales of tens of seconds or several minutes.
Therefore, we used images obtained from the SDO/AIA observations, which have a higher time resolution. 
The SDO/AIA instrument takes full-sun images of nine UV and EUV broadband channels. 
Among these channels, we chose 211~\AA~, which is responsive to the emission of $10^{6.3}$ K plasma \citep{Boerner2012}. 
The spatial and time resolutions of each filter were $0.6\arcsec$ and 12 s, respectively. 
AIA images were calibrated by the \verb|aia_prep| routine in SSW to eliminate instrumental effects. 

\section{Event Detection} \label{sec:detect}
\begin{figure}[tp]
	\centering
	\includegraphics[width=1\linewidth]{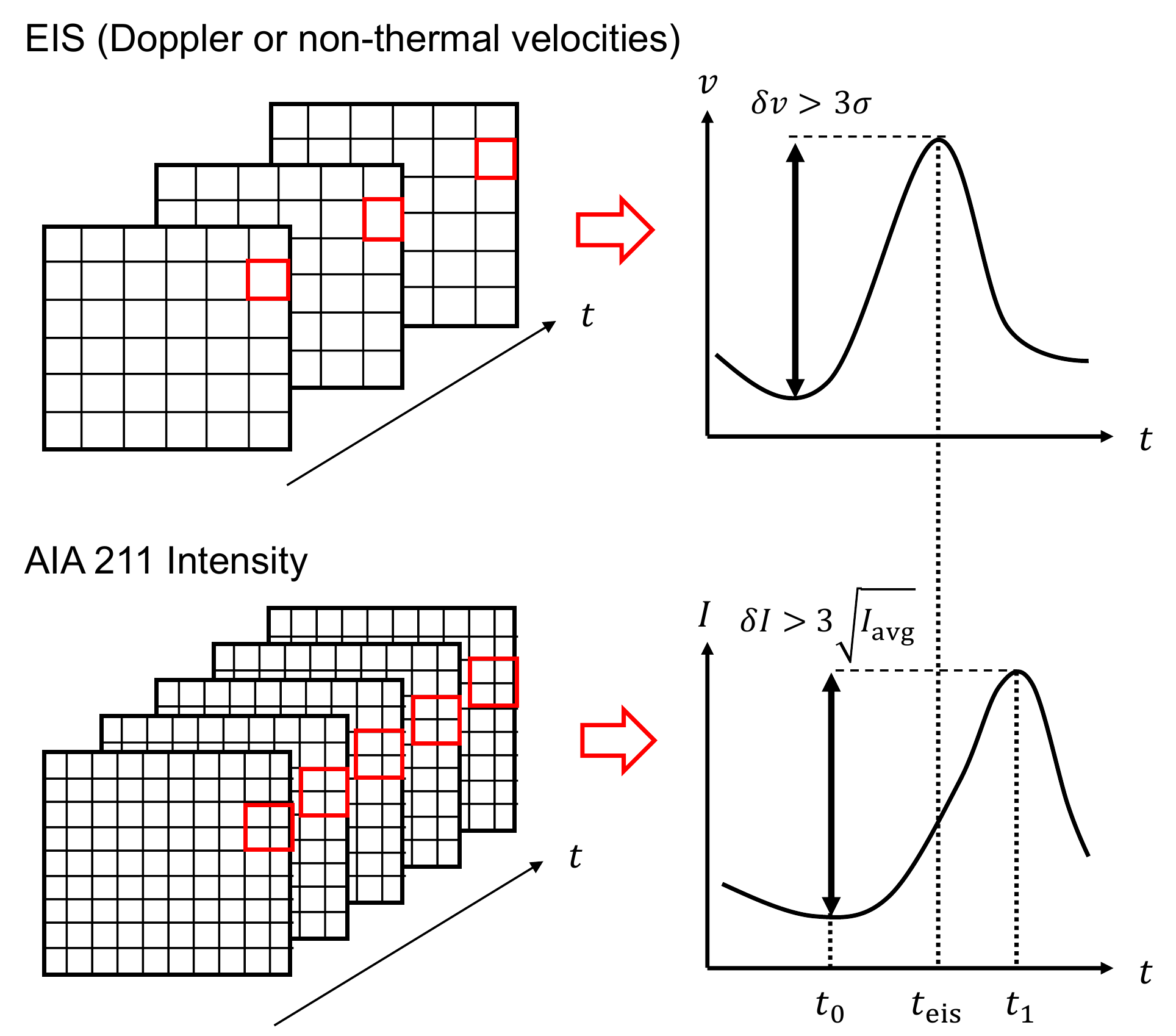}
	\caption{Flow of the detection of EIS Doppler/non-thermal velocity enhancements. First, we detect a Doppler/non-thermal velocity enhancement that exceeds the threshold. Second, an intensity enhancement of the AIA 211~\AA~light curve around the velocity enhancement is detected if it exists. Finally, only velocity enhancements with peaks ($t_\mathrm{eis}$) between $t_0$ and $t_1$ are selected as valid events for later analysis. }
	\label{fig:scheme}
\end{figure}

Figure~\ref{fig:scheme} presents a schematic diagram of the flow of the detection of the EIS Doppler/non-thermal velocity enhancements. 
We assumed that small-scale flares are enhancements of either Doppler or non-thermal velocities defined by the following equation:
 \begin{equation}
\delta v > 3 \times  \sigma
\end{equation}
where $\delta v$ and $\sigma$ represent changes in the absolute value of the Doppler/non-thermal velocity and the standard deviation for 10 min before the beginning of the enhancement, respectively. 
Sometimes, Doppler and non-thermal velocities are not derived accurately because of the lack of intensity, especially in Fe\,{\sc xiv} and \,{\sc xvi}. 
Therefore, we neglected the pixels that had an intensity less than $5 \times 10^4~\mathrm{erg~s^{-1}~cm^{-2}~sr^{-1}}$ as noise. 
We decided this threshold based on the median of a Fe \,{\sc xvi} intensity map shown in Figure~\ref{fig:eismap} ($\simeq 4.9 \times 10^4~\mathrm{erg~s^{-1}~cm^{-2}~sr^{-1}}$) because the Doppler velocity and line width in about half the area of the map are unstable. 
After the detection of enhancements, we obtained the solar-X and -Y coordinates and the scan time of each enhancement. 
When peaks of multiple enhancements appear at adjacent pixels simultaneously, they are regarded as a single event. 
In this case, the event coordinate and energy are defined as the geometric center of gravity and their sum in each pixel, respectively. 
We derived the occurrence time of each enhancement using solar-X and the scan step time because small events are thought to be sensitive to time resolution. 
The EIS raster scan begins from the west side and requires approximately 4.6 s for each scan step (pointing position). 
This detection was performed for Fe\,{\sc xiv} (264.92~\AA), \,{\sc xv} (284.20~\AA), \,{\sc xvi} (262.98~\AA) images.

We obtained SDO/AIA light curves around energy enhancements detected by EIS observations. 
However, it is necessary to align both the coordinates of the Hinode/EIS and SDO/AIA data. 
We applied a template matching method between a pair of EIS and AIA images. 
First, an EIS raster image was converted to have the same spatial resolution as the AIA using bilinear interpolation. 
Second, we rigidly slide the converted EIS intensity image as a template over the AIA image and calculate the square root of difference in intensities between them at each location. 
Finally, we derive the spatial offset from where the difference becomes the minimum.
We used the EIS Fe\,{\sc xiv} (264.92~\AA) and the AIA 211~\AA image observed at 18:12:04 UT and 18:11:59 on November 09 for the matching. 
The intensities in both snapshots are normalized to 0 -- 1 range because the response is different from each other. 
Then, we applied the spatial offset found as described above to all other EIS images.

\begin{figure*}[tp]
	\centering
	\includegraphics[width=1\linewidth]{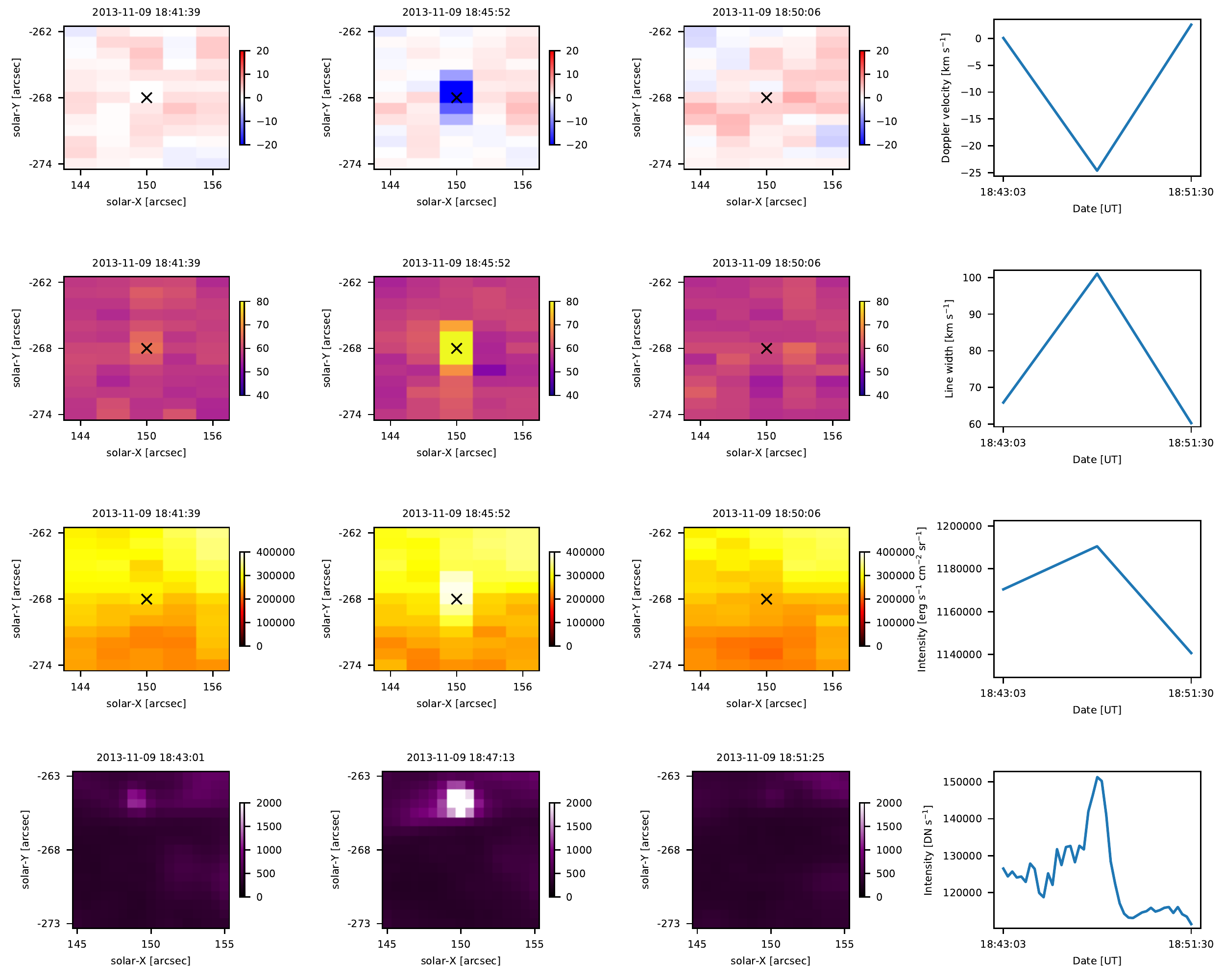}
	\caption{An example of time series of the Fe\,{\sc xv} Doppler velocity (first row), line width (second row), EIS intensity (third row), and AIA 211~\AA~intensity (fourth row) around a detected enhancement. The first and third columns represent the maps before and after enhancement. The maps in the second column are those at the peak of the Doppler velocity enhancement. The right column shows a temporal series of the velocities at the center (black ticks) of the map and the light curves around the enhancement obtained by EIS and AIA. }
	\label{fig:sample}
\end{figure*}

After the EIS-AIA calibration, we obtained an AIA light curve around the EIS velocity enhancement for 150~s before and after the occurrence. 
We used this time range because small-scale heating events such as microflares and nanoflares are thought to have time scales of tens of seconds or several minutes. 
The AIA lightcurve is obtained by integrating over $16 \times 16$ pixels centered on the calibrated location of the EIS enhancement.
This area is somewhat large when compared to the smallest event; however, we used this value by considering the inaccuracy of the estimated EIS-AIA spatial offset. 
In fact, as we will show later (Figure~\ref{fig:sample}), the locations of EIS and AIA enhancements are sometimes misaligned. 
Moreover, this area might be smaller than the largest AIA enhancement. 
However, even if the entire AIA brightening does not fit the lightcurve area, it does not affect our quantitative analysis; while we need the AIA lightcurve for the event selection, the Doppler velocities, non-thermal, and thermal energies all needed to derive the energy conversion rate use solely the EIS data (see next section).
We detected enhancements in the AIA 211~\AA~light curve accompanying EIS energy enhancement. 
An AIA enhancement begins when the time difference of intensity exceeds the threshold $3 \sqrt{I_\mathrm{avg}}$, where $I_\mathrm{avg}$ represents the square root of the average intensity 1 min before the beginning of the EIS enhancement. 
Sometimes, multiple AIA enhancements existed for single EIS enhancements; therefore, we chose the AIA enhancement that begins at the time closest to the EIS event. 
When there were no AIA enhancements that satisfy the condition above, EIS enhancement was not employed for the analysis below. 

We selected EIS velocity enhancements that occur with AIA enhancement to compare the energy balance between them. 
We derived the relative time of the EIS energy enhancement to an AIA light curve enhancement $t_\mathrm{rel}$ using the following equation: 
 \begin{equation}
t_\mathrm{rel} = \frac{t_\mathrm{eis} - t_0}{t_1 - t_0}
\end{equation}
where $t_\mathrm{eis}$, $t_0$, and $t_1$ represent the time when the EIS velocity enhancement is detected, and the beginning and peak of the AIA enhancement. 

Figure~\ref{fig:sample} presents an example of temporal series of the Fe\,{\sc xv} Doppler velocity, line width, and intensity and AIA 211~\AA~maps. 
The first and second rows show the Doppler velocity and line width maps around the detected enhancement and the temporal series at the center (black ticks) of the maps. 
The third and fourth rows represent the time series of the intensity maps and light curves around the velocity enhancement obtained from the EIS and AIA, respectively. 
Comparing the panels in the second column, the location of an event seems slightly misaligned between the EIS and AIA (the AIA enhancement has a larger solar-Y). 
This gap is caused by the inaccuracy of our alignment; however, this misalignment is compensated for by obtaining the AIA light curve over a somewhat large area ($16 \times 16$) to avoid missing events.

\section{Energy Estimation} \label{sec:energy}
To determine the Doppler motion, non-thermal, and thermal energies, we estimated the electron density using the \verb|eis_density| function in SSW. 
This function returns the electron number density as a response to the input of a pair of spectra in the Fe\,{\sc xiv} windows. 
Sometimes, the density derived by this method does not have a valid number owing to photon noise or lack of data. 
Pixels with these errors are neglected in later analysis.

The Doppler motion energy in each EIS pixel is defined as follows: 
\begin{equation}
E_\mathrm{Dop} = \frac{1}{2}m n_e v_\mathrm{Dop}^2 S_\mathrm{eis}^\frac{3}{2}
\label{eq:dop}
\end{equation}
where $m$ represents the mean particle mass, and its value is $2.08 \times 10^{-24}$ g in this study. 
$n_e$ and $S_\mathrm{eis}$ represent the electron number density estimated by the \verb|eis_density| function and 1 pixel area of the EIS image, respectively. 
Similarly, non-thermal energy is defined as
\begin{equation}
E_\mathrm{nth} = \frac{1}{2}m n_e v_\mathrm{nth}^2 S_\mathrm{eis}^\frac{3}{2}. 
\label{eq:nth}
\end{equation} 
When multiple enhancements are detected in adjacent pixels at the same time, the summed energy is multiplied by $\sqrt{N}$, the number of pixels, to correct the volume.  

We compared the Doppler motion and non-thermal energies at the peak of each enhancement and the difference in thermal energy between the peak and the background. 
We define the indicator of the energy balance $\phi_\mathrm{Dop}$ and $\phi_\mathrm{nth}$ using the following equation:
 \begin{eqnarray}
\phi_\mathrm{Dop} = \frac{\Delta E_\mathrm{th}}{E_\mathrm{Dop}} \\
\phi_\mathrm{nth} = \frac{\Delta E_\mathrm{th}}{E_\mathrm{nth}}
\label{eq:phi}
\end{eqnarray}
where $\Delta E_\mathrm{th}$, $E_\mathrm{Dop}$, and $E_\mathrm{Dop}$ represent the changes in thermal energy, Doppler motion, and non-thermal energies, respectively. 
This difference of thermal energy is estimated as follows: 
 \begin{equation}
\Delta E_\mathrm{th} = 3 k_B n_{e1} T_1 S_\mathrm{eis}^\frac{3}{2} - 3 k_B n_{e0} T_0 S_\mathrm{eis}^\frac{3}{2}
\label{eq:eth}
\end{equation}
where subscripts 0 and 1 indicate the background and peak, respectively. 
We assume that the background temperature $T_0$ and density $n_{e0}$ are medians of those estimated by the entire EIS observations. 
We did not use pre-event temperature and density for background values because the time scale of transient brightenings or nanoflares (from tens of seconds to several minutes) is shorter than that of EIS observations ($\ge 254~\mathrm{s}$). 

\section{Result} \label{sec:result}
Based on the criteria in Section~\ref{sec:detect}, the numbers (occurrence frequencies [$\mathrm{s^{-1} cm^{-2}}$]) of the detected Doppler velocity enhancements with AIA enhancements are 6077 ($1.5 \times 10^{-21}$) for Fe\,{\sc xiv}, 10727 ($2.7 \times 10^{-21}$) for Fe\,{\sc xv}, and 4183 ($1.0 \times 10^{-21}$) for Fe\,{\sc xvi}. 
Those of the non-thermal velocity are 583 ($1.4 \times 10^{-22}$) , 889 ($2.2 \times 10^{-22}$),  and 1 ($2.5 \times 10^{-25}$) for Fe\,{\sc xiv}, Fe\,{\sc xv}, and Fe\,{\sc xvi}, respectively. 

\begin{figure*}[tp]
	\centering
	\includegraphics[width=1\linewidth]{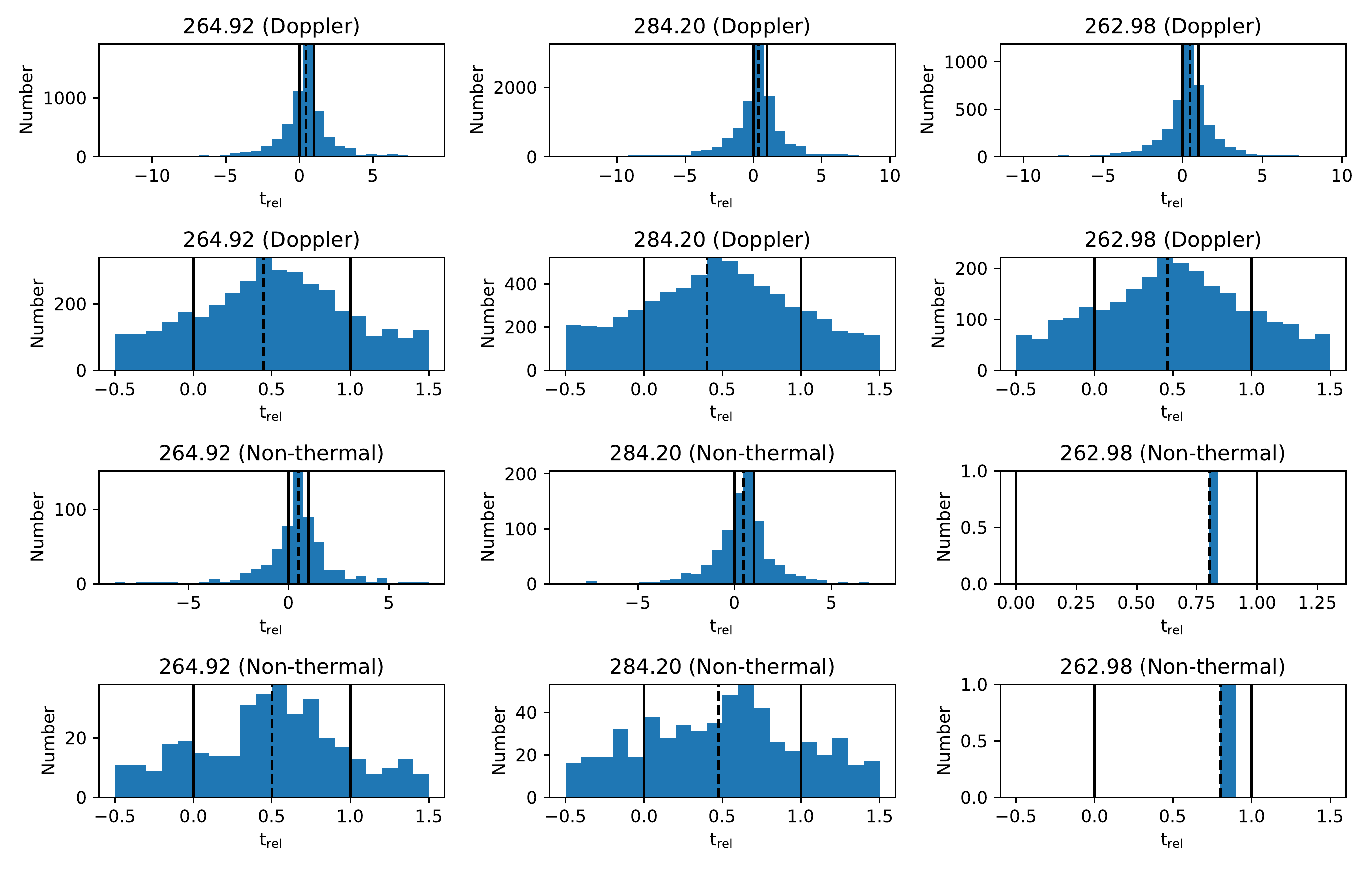}
	\caption{The first and third rows represent the entire $t_\mathrm{rel}$ distributions of energy enhancements. The second and fourth rows show the detailed distributions of the first and third rows in the range of  $-0.5 \le t_\mathrm{rel} \le 1.5$. The solid lines indicate $t_\mathrm{rel} = 0$ and $t_\mathrm{rel} = 1$. The dashed line in each panel indicates the median of $t_\mathrm{rel}$. }
	\label{fig:rel}
\end{figure*}
The first and third rows in Figure~\ref{fig:rel} present the entire distributions of the relative occurrence time of EIS energy enhancements to AIA enhancements, $t_\mathrm{rel}$, detected using Doppler and non-thermal velocities, respectively. 
The second and fourth rows represent the detailed distributions of $t_\mathrm{rel}$ between -0.5 \-- 1.5. 
The solid lines indicate $t_\mathrm{rel} = 0$ and $t_\mathrm{rel} = 1$, which indicate the beginning and peak of the AIA enhancement, respectively. 
The dashed line in each panel in Figure~\ref{fig:rel} indicates the median of $t_\mathrm{rel}$ for each wavelength. 
We find that around 40\% of the velocity enhancements occur between $t_\mathrm{rel} = 0$ and $t_\mathrm{rel} = 1$. 
The median $t_\mathrm{rel}$ is 0.4 -- 0.5 for each channel except for the Doppler velocity enhancements detected by Fe\,{\sc xvi}.
According to some loop heating simulations \citep[{\it e.g.,\rm}][]{Bowness2013, Botha2011}, the released energy is converted into kinetic energy just after the beginning of an enhancement, prior to $t_{\rm rel} \approx 0.5$ after which it is transformed into thermal energy. 
Accordingly, $t_\mathrm{rel}$ should distribute just after the beginning (for example, $t_\mathrm{rel} \simeq 0.1$).
The distribution of $t_\mathrm{rel}$ shown here does not directly support this process.

\begin{figure*}[tp]
	\centering
	\includegraphics[width=1\linewidth]{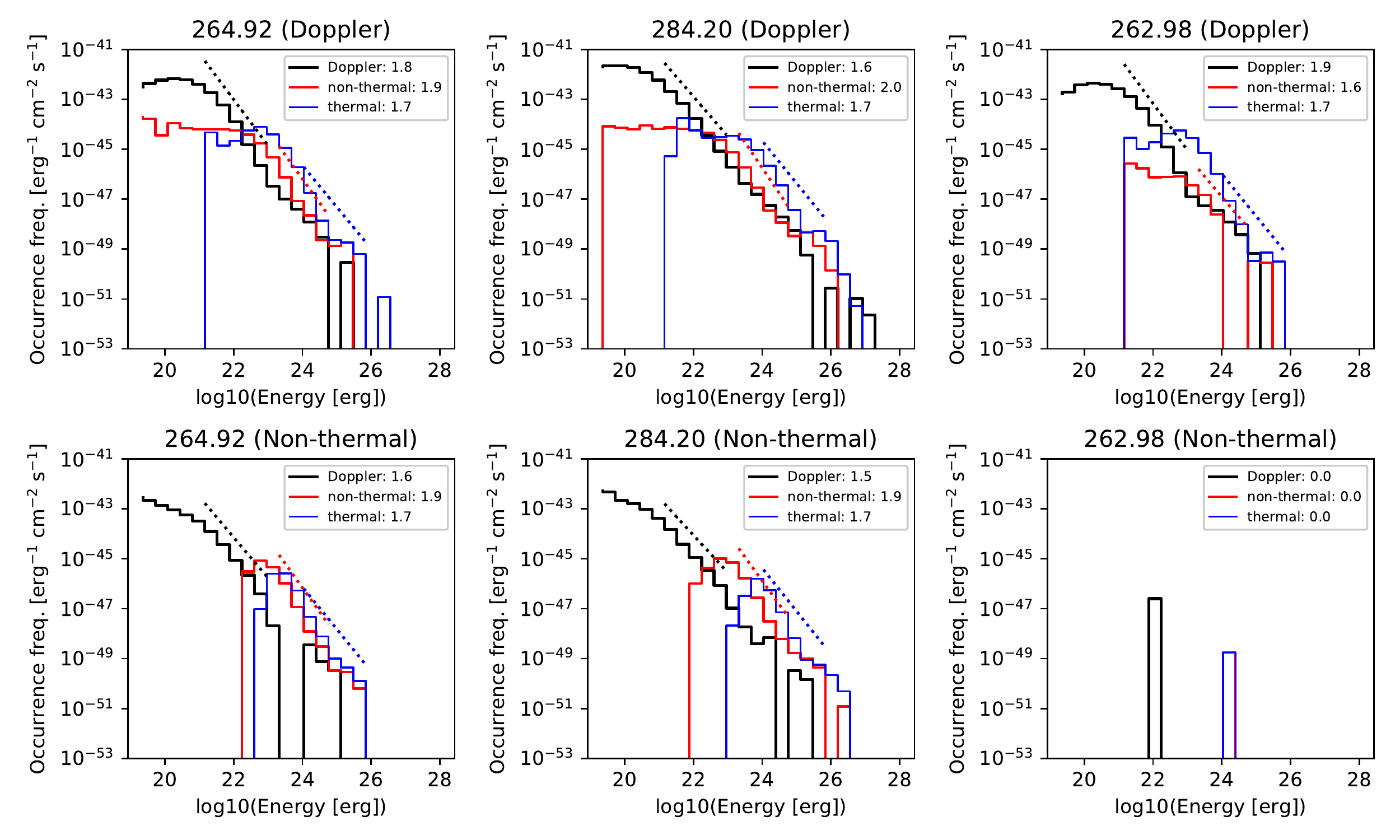}
	\caption{Occurrence frequency distributions of enhancements detected by Fe\,{\sc xiv} ({\it left}), Fe\,{\sc xv} ({\it center}), and Fe\,{\sc xvi} ({\it right}). The top and bottom panels represent the distributions detected using Doppler and non-thermal velocities, respectively. The black, red, and blue solid lines represent distributions of Doppler motion, non-thermal energy, and thermal energy, respectively. The dotted lines indicate power-law fitted lines in the energy range of $10^{21} \le E \le 10^{23}~\mathrm{[erg]}$ (Doppler), $10^{23} \le E \le 10^{25}~\mathrm{[erg]}$ (non-thermal), and $10^{24} \le E \le 10^{26}~\mathrm{[erg]}$ (thermal). The numbers in the legend represent the power-law indices.}
	\label{fig:ene}
\end{figure*}
Figure~\ref{fig:ene} presents the occurrence frequency distributions as a function of energy derived by Fe\,{\sc xiv} (left), Fe\,{\sc xv} (center), and Fe\,{\sc xvi} (right). 
The top and bottom panels represent the distributions detected using Doppler and non-thermal velocities, respectively. 
The black, red, and blue lines represent the distributions of Doppler motion, non-thermal energy, and thermal energy, respectively. 
The dotted lines indicate power-law fitted lines in the energy range of $10^{21} \le E \le 10^{23}~\mathrm{[erg]}$ (Doppler), $10^{23} \le E \le 10^{25}~\mathrm{[erg]}$ (non-thermal), and $10^{24} \le E \le 10^{26}~\mathrm{[erg]}$ (thermal). 
The numbers in the legend represent the power-law indices ($\alpha$ in Equation~\ref{eq:alpha}). 
We chose the above energy ranges to roughly maximize the indices; however, there are no distributions which have a power-law index greater than two, {\it i.e.,} the results do not meet the condition necessary for nanoflare heating to be dominant. 

\begin{figure*}[tp]
	\centering
	\includegraphics[width=1\linewidth]{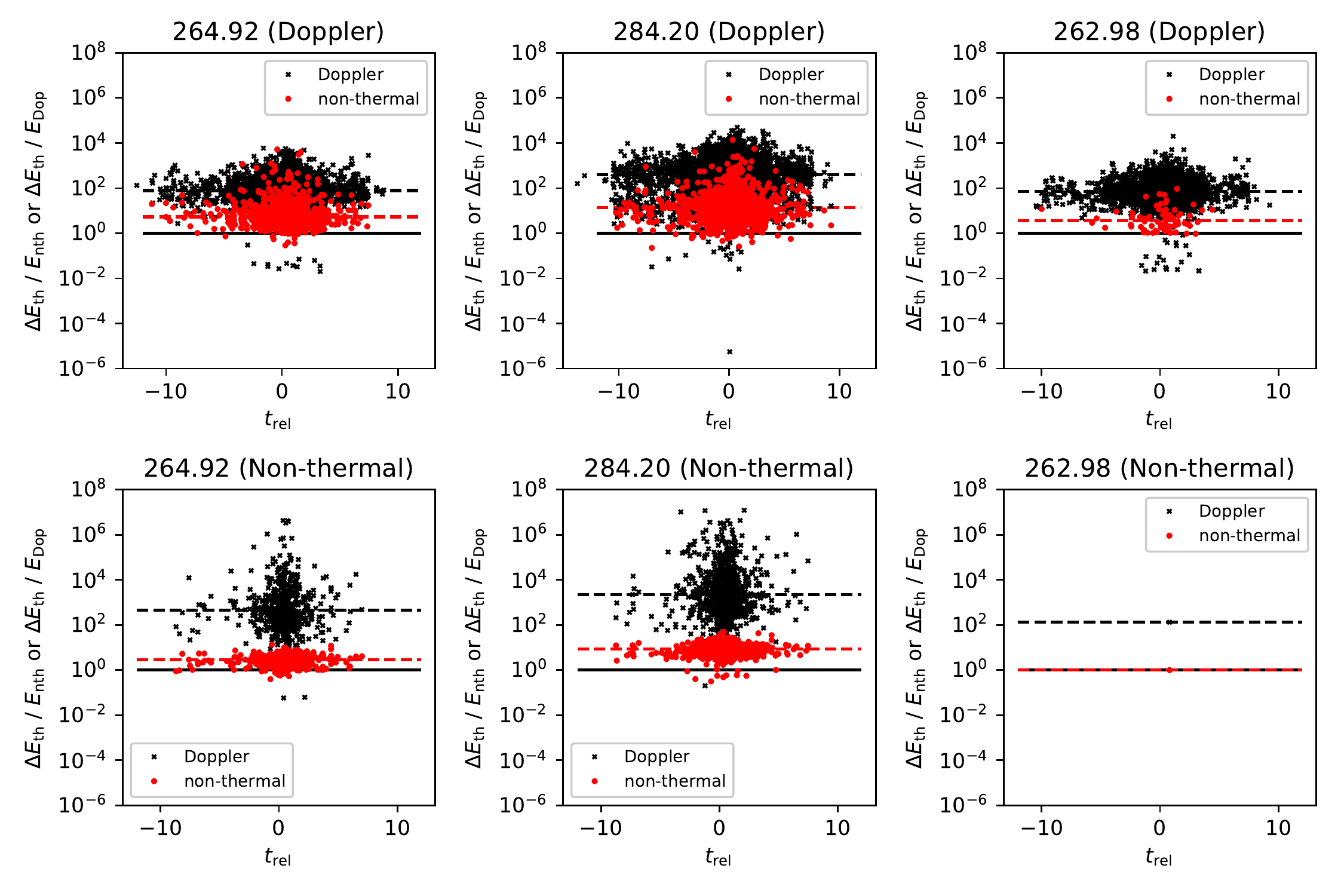}
	\caption{Relative times of an EIS energy enhancement to an AIA enhancement $t_\mathrm{rel}$ vs energy ratio $\phi_\mathrm{Dop}$ and $\phi_\mathrm{nth}$ estimated by Fe\,{\sc xiv} ({\it left}), Fe\,{\sc xv} ({\it center}), and Fe\,{\sc xvi} ({\it right}). The top and bottom panels present the distributions of Doppler and non-thermal velocity enhancements, respectively. Black and red circles represent the distributions of $\phi_\mathrm{Dop}$ and $\phi_\mathrm{nth}$, respectively. The horizontal solid line indicates the location $\phi=0$. The black and red dashed lines indicate the medians of $\phi_\mathrm{Dop}$ and $\phi_\mathrm{nth}$, respectively.}
	\label{fig:vs}
\end{figure*}
Figure~\ref{fig:vs} presents the associations between $t_\mathrm{rel}$ and the energy ratio $\phi_\mathrm{Dop}$ and $\phi_\mathrm{nth}$. 
The top and bottom panels show enhancements detected using Doppler and non-thermal velocities, respectively. 
Black ticks and red circles represent the scatter plot of $\phi_\mathrm{Dop}$ and $\phi_\mathrm{nth}$.  
The horizontal solid line in each panel indicates where $\phi=1$, which implies that the released thermal and Doppler motion (non-thermal) energies are balanced. 
The black and red dashed lines represent the medians of $\phi_\mathrm{Dop}$ and $\phi_\mathrm{nth}$, respectively. 
According to the medians, the released Doppler motion energy is 0.1 -- 1\% of the change in thermal energy. 
In addition, the released non-thermal energy is approximately 10 -- 100\% of the difference in thermal energy. 
There are no clear correlations between $t_\mathrm{rel}$ and $\phi$. 

\section{Energy Balance Analysis} \label{sec:balance}
In this section, we estimate the energy balance between the energy losses by radiation and thermal conduction, and the heat flux inputs, for the detected enhancements.. 
First, we derive the radiative and conductive energy loss of the observed active region. 
Assuming the active region as a half-sphere, the radius $r$ can be described as $r=\sqrt{S/\pi}$, where $S=8.7 \times 10^{19}~\mathrm{cm^2}$ represents the EIS observation area. 
First, radiative loss flux is estimated as follows:
 \begin{eqnarray}
 F_r = n_e^2P(T)VS^{-1} \\
 P(T) = 10^{-17.73}T^{-\frac{2}{3}} \\
 V=\frac{2}{3}\pi r^3
 \end{eqnarray}
where $n_e=1.8 \times 10^{9}~\mathrm{cm^{-3}}$ and $T=2.7 \times 10^6~\mathrm{K}$ represent the medians of electron density estimated by the \verb|eis_density| function and temperature, respectively. 
The mean temperature was derived by comparing the observed ratio of Fe\,{\sc xiv} and Fe\,{\sc xvi} with that of the CHIANTI database, although the estimated temperature depends on the selected ion pair. 
We used medians of density and temperature instead of their average values because these parameters are highly sensitive to the observational noise, especially in the case of dense plasma. 
$P(T)$ is the radiative loss function when $10^{6.3}<T<10^{7.0}~\mathrm{K}$~\citep{Raymond1976, Rosner1978}. 
In contrast, conduction loss flux is derived as follows:
 \begin{equation}
F_c = 2 \kappa T^\frac{7}{2} \pi r S^{-1}
 \end{equation}
where $\kappa = 1.1 \times 10^{-6}~\mathrm{erg~cm^{-3}~s^{-1}~K^\frac{7}{2}}$ \citep{Spitzer1956}.
Consequently, the radiative and conductive loss fluxes are $1.1 \times 10^6~\mathrm{erg~s^{-1}~cm^{-2}}$ and $1.4 \times 10^7~\mathrm{erg~s^{-1}~cm^{-2}}$, respectively. 
Therefore, the total loss flux is approximately $1.5 \times 10^7~\mathrm{erg~s^{-1}~cm^{-2}}$. 
This value is consistent with the result of~\citet{Withbroe1977} ($\simeq 10^7~\mathrm{erg~s^{-1}~cm^{-2}}$).

Second, we estimated the mean heating flux from AIA 211~\AA~intensity enhancements. 
We derived the flux using 1 h AIA observation maps similar to Figure~\ref{fig:211} from 18:07:13 UT on November 09. 
From the light curve of each $4 \times 4$ macro-pixels, an AIA enhancement begins when the increase in intensity exceeds the threshold $3 \sigma$, where $\sigma$ represents the 1 min mean intensity before the beginning of the enhancement. 
The events in adjacent pixels that have any overlap in the rising phase are regarded as a single event. 
The event area is calculated from the number of pixels where the enhancement spans.
The released energy of each enhancement is defined as the change in the amount of thermal energy, similar to Equation~\ref{eq:eth}. 
Electron densities $n_{e0}$ and $n_{e1}$ are defined as follows:
 \begin{eqnarray}
n_e = \sqrt{\frac{\mathrm{EM}}{S_\mathrm{aia}^\frac{3}{2}}} \\
\mathrm{EM} = \frac{I}{F(T)}
\end{eqnarray}
where $\mathrm{EM}$, $I$, $S_\mathrm{aia}$, and $F(T)$ represent the emission measure, intensity of AIA 211~\AA, event area, and response of AIA 211~\AA~ filter to plasma at temperature $T$ \citep{Boerner2012}. 
The filling factor is assumed to be unity in this study. 
We assumed that $T_0$ and $T_1$ to be $1~\mathrm{MK}$ and $5~\mathrm{MK}$ similar to the method of \citet{Shimizu1995}.
We avoid using the EIS density and temperature for this calculation because the temporal resolution of the EIS images is about 20 times longer than that of AIA. 
Moreover, there are many AIA enhancements which have a time scale shorter than the cadence of EIS observations as we will show later (Figure~\ref{fig:durs}).
Thus, the number of detected events is approximately $1.6 \times 10^4$ with the occurrence frequency of $1.2 \times 10^{-21}~\mathrm{s^{-1} cm^{-2}}$. 
The mean flux of the detected enhancements is approximately $1.2 \times 10^5~\mathrm{erg~s^{-1}~cm^{-2}}$. 
Based on our results, the Doppler kinetic and non-thermal energies of AIA enhancements can be roughly estimated as $1.2 \times 10^2$ -- $1.2 \times 10^3~\mathrm{erg~s^{-1}~cm^{-2}}$ and $1.2 \times 10^4$ -- $1.2 \times 10^5~\mathrm{erg~s^{-1}~cm^{-2}}$, respectively. 
Therefore, the total AIA energy flux is approximately $1.3 \times 10^5$ -- $2.4 \times 10^5~\mathrm{erg~s^{-1}~cm^{-2}}$, which is approximately 1 -- 2\% of the sum of the conduction and radiative loss fluxes. 

\begin{table*}[tpb!]
	\centering
	\caption{Energy flux of EIS Doppler/non-thermal velocity enhancements [$\mathrm{erg~s^{-1} cm^{-2}}$] and contribution to the heating [\%]}
	  \begin{tabular}{rrrr|r}
	  Ion species & Doppler flux & Non-thermal flux & Thermal flux & Required flux \\ \hline
Fe\,{\sc xiv} (Doppler) & $2.9\times 10^{1}$ ($10^{-4}$\%) & $1.0\times 10^{2}$ ($10^{-3}$\%) & $6.9 \times 10^2$ ($10^{-2}$\%) & $1.5 \times 10^7$  (100\%)\\
Fe\,{\sc xiv} (non-thermal) & $4.2\times10^{0}$ ($10^{-5}$\%) & $5.4\times 10^{1}$ ($10^{-4}$\%) & $1.1 \times 10^{2}$ ($10^{-3}$\%) &  \\
Fe\,{\sc xv} (Doppler) & $7.0\times 10^{2}$ ($10^{-2}$\%) & $6.0 \times 10^{2}$ ($10^{-3}$\%) & $5.5 \times 10^{3}$ ($10^{-2}$\%) &  \\
Fe\,{\sc xv} (non-thermal) & $1.1\times 10^{1}$ ($10^{-4}$\%) & $2.2 \times 10^{2}$ ($10^{-3}$\%) & $8.6 \times 10^{2}$ ($10^{-2}$\%) & \\
Fe\,{\sc xvi} (Doppler) & $1.8 \times 10^{1}$ ($10^{-4}$\%) & $1.5\times 10^{1}$ ($10^{-4}$\%) & $2.9 \times 10^{2}$ ($10^{-3}$\%) & \\
Fe\,{\sc xvi} (non-thermal) & $2.8 \times 10^{-3}$ ($10^{-8}$\%) & $3.7\times 10^{-1}$ ($10^{-6}$\%) & $3.7 \times 10^{-1}$ ($10^{-5}$\%) & \\
	   \end{tabular}
	\label{tab:flux}
\end{table*}

We calculated the energy flux of the detected EIS Doppler motion and non-thermal enhancements accompanied by the AIA intensity enhancement as follows:
 \begin{equation}
F_i = \sum E  S^{-1} \tau^{-1}
 \end{equation}
where $E$ represents the estimated Doppler, non-thermal or thermal energies of each enhancement. 
The observation duration $\tau$ is 30,831 s, as described in Section~\ref{sec:obs}. 
Table~\ref{tab:flux} shows Doppler motion, non-thermal and thermal energy fluxes detected using Doopler/non-thermal velocity by each ion channel and their contribution to the active region heatnig. 
The fluxes depend on the wavelengths and velocities; however, the detected input flux is much less than the loss flux in any case. 
The ratio of contributions are roughly $F_\mathrm{Dop}:F_\mathrm{nth}:F_\mathrm{th}=1:10:100$. 

\section{Discussion and Summary} \label{sec:summary}
In this study, we estimated the ratio of the difference in thermal energy $\Delta E_\mathrm{th}$ to $E_\mathrm{Dop}$ based on Doppler motion and $E_\mathrm{nth}$  based on non-thermal velocities using Hinode/EIS Fe\,{\sc xiv}, Fe\,{\sc xv}, and Fe\,{\sc xvi} spectroscopic observations. 
As a result, $E_\mathrm{Dop}$ and $E_\mathrm{nth}$ tend to be 0.1 -- 1\% and 10 -- 100\% of $\Delta E_\mathrm{th}$ in the typical enhancements detected by all three wavelengths, respectively. 
The contribution of the energy fluxes of detected EIS velocity enhancements was less than 0.1\% of the estimated total losses due to radiation and conduction for all three ion channels. 
Moreover, using the energy conversion rate (Section~\ref{sec:result}), we estimated the contribution of AIA transient brightenings to active region heating to be at most 2\% of these losses. 
This shortage is probably caused by the lack of detection and/or the contribution of other processes. 

\begin{figure}[tp]
	\centering
	\includegraphics[width=1\linewidth]{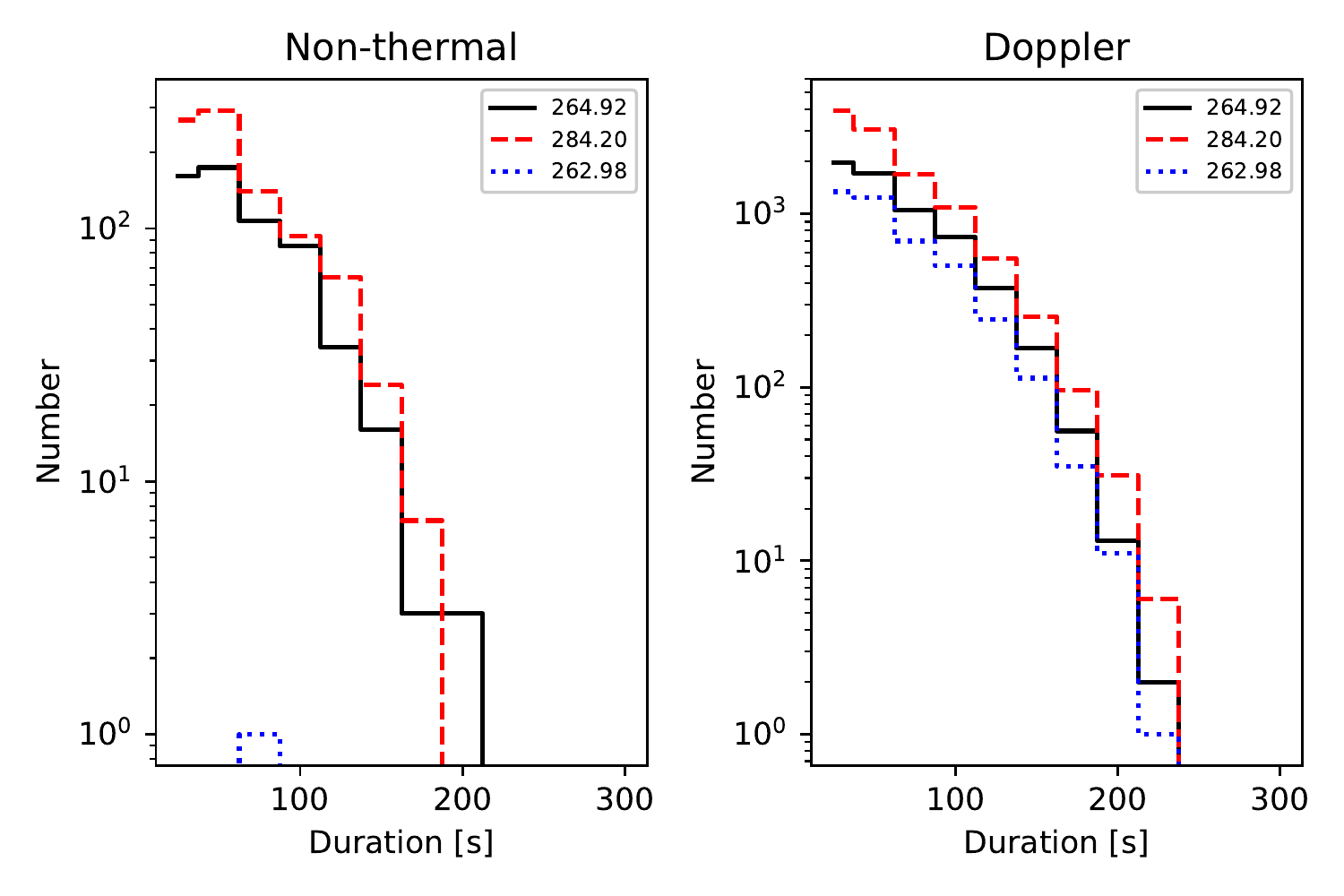}
	\caption{Distribution of durations from beginnings of AIA enhancements to their peaks, which have an EIS velocity enhancement between them. The left (right) panel shows the distribution of Doppler (non-thermal) velocity. The black, red, and blue dotted lines represent the histograms of Fe\,{\sc xiv}, Fe\,{\sc xv}, and Fe\,{\sc xvi}, respectively. }
	\label{fig:durs}
\end{figure}
Each histogram in Figure~\ref{fig:durs} represents a distribution of durations from the beginning of AIA brightening to its peak accompanying EIS velocity enhancement. 
The Black solid, red dashed, and blue dotted lines represent distributions derived from Fe\,{\sc xiv}, Fe\,{\sc xv}, and Fe\,{\sc xvi}, respectively. 
The medians of durations derived from these distributions are approximately 48 s. 
We estimated the required cadence of EUV spectroscopic imaging observations to detect enhancements of Doppler and non-thermal velocities. 
As described in Section~\ref{sec:result}, the median of $t_\mathrm{rel}$ is approximately 0.4. 
Then, the time scale of medium EIS velocity enhancements is roughly estimated as $48 \times 0.4 \simeq 19$ s, whereas the cadence of Hinode/EIS data used in this study is 254 s. 
The next Japanese satellite mission, Solar-C/EUV High-throughput Spectroscopic Telescope (EUVST), will achieve a temporal resolution that is approximately 10 times higher than that of Hinode/EIS. 
Therefore, the EUVST will provide a more accurate process of energy release during these transient brightenings after its launch in mid 2020. 

According to the distributions in Figure~\ref{fig:ene}, the smallest released energy is much smaller than that of previous observations. 
However, the Doppler velocity is corrected using the median of each snapshot. 
This correction might cause underestimation of the Doppler velocity. 
In contrast, the Doppler and non-thermal energies estimated by Fe\,{\sc xvi} are much larger than those of the other channels. 
This is probably because many pixels in this channel have spectra that do not have an adequate photon count to derive the velocities accurately, although we neglected darker pixels in the detection criteria. 

We used some threshold for our detection criteria to avoid detecting noise as an event. 
Definitely, the looser the thresholds to use, the greater the number of events. 
However, as shown in Figure~\ref{fig:ene}, the power-law indices are less than two in all occurrence frequency distributions.
Therefore, the energy contribution of smaller events than those we detected in this study is not a dominant if we assume that the power-law index does not change even in smaller energy range.

There are some neglected processes and influential assumptions in our analysis. 
\citet{Schmelz2001} and \citet{Warren2008} revealed that a coronal loop contains plasma that has a wide temperature range, which implies that the loop consists of finer tubes. 
The volumetric filling factor of the loop was estimated to be approximately $10\%$ by \citet{Warren2008}. 
In contrast, \citet{Sakamoto2009} showed that the volumetric filling factors for {\it SXT} and the {\it Transition Region And Coronal Explorer }~\citep[TRACE:][]{Handy1999} loops are $\simeq 2\%$ and $\simeq 70\%$, respectively. 
These results suggest that the filling factor depends on the observational wavelength.  
Accordingly, the input energy flux might be overestimated by an order of magnitude because we assume the filling factor as unity to estimate the AIA thermal energy.
Moreover, we assumed that the ion temperature is the same as that of the electrons; however, \citet{Imada2009} reported that they have different temperatures, especially in an active region core.  
Non-equilibrium ionization (NEI) also affects the detection and energy estimation of impulsive heating events~\citep{Orlando1999, Reale2008, Imada2011a}. 
As for energy balance analysis, heating input from dark jets, those events showing Doppler shift but no enhancements in AIA images as reported by \citet{Young2015}, are neglected. 
The energy contribution of these events will be derived in the future works.

The authors thank K. Kusano, K.D. Leka, and P. Sung-Hong for fruitful discussions.
This work was partially supported by a Grant-in-Aid for 17K14401, 15H05816, and JSPS Fellows, and the Program for Leading Graduate Schools, ``PhD Professional: Gateway to Success in Frontier Asia'' by the Ministry of Education, Culture, Sports, Science and Technology. 
Hinode is a Japanese mission developed and launched by ISAS/JAXA, in collaboration with NAOJ as a domestic partner and NASA and STFC (UK) as international partners. 
Scientific operation of the Hinode mission was conducted by the Hinode science team organized at ISAS/JAXA. 
This team mainly consists of scientists from institutes in partner countries. 
Support for the post-launch operation is provided by JAXA and NAOJ (Japan), STFC (U.K.), NASA (U.S.A.), ESA, and NSC (Norway). 
The {\it Solar Dynamics Observatory} is a part of NASA's Living with a Star program. 
A part of this study was carried out using the computational resources of the Center for Integrated Data Science, Institute for Space-Earth Environmental Research, Nagoya University.


\bibliography{reference}{}
\bibliographystyle{aasjournal}



\end{document}